\documentclass[prb,twocolumn,showpacs,preprintnumbers,amsmath,amssymb,10pt]{revtex4-1}

\usepackage{color}
\definecolor{blau}{rgb}{0,0,1}

\usepackage[T1]{fontenc}
\usepackage[pdftex]{graphicx}
\usepackage{dcolumn}
\usepackage{bm}

\usepackage{mathcomp}

\DeclareGraphicsExtensions{.pdf}   
\DeclareMathOperator{\e}{e}

\usepackage[justification=raggedright,singlelinecheck=false]{caption}

\begin{document}

\title{Hybrid functional calculations of the Al impurity in $\alpha$-quartz: Hole localization and electron paramagnetic resonance parameters}
\author{Roland Gillen}\email{rg403@cam.ac.uk}
\author{John Robertson}
\affiliation{Department of Engineering, University of Cambridge, Cambridge CB3 0FA, United Kingdom}

\date{\today}

\begin{abstract}
We performed first-principle calculations based on the supercell and cluster approaches to investigate the neutral Al impurity in smoky quartz. Electron paramagnetic resonance measurements suggest that the oxygens around the Al center undergo a polaronic distortion which localizes the hole being on one of the four oxygen atoms. We find that the screened exchange hybrid functional successfully describes this localization and improves on standard local density approaches or on hybrid functionals that do not include enough exact exchange such as B3LYP. We find a defect level at about 2.5 eV above the valence band maximum, corresponding to a localized hole in a O 2p orbital. The calculated values of the g tensor and the hyperfine splittings are in excellent agreement with experiment. 
\end{abstract}


\maketitle

\section{Introduction}
Silicon dioxide is among the most commonly encountered substances in both daily life and in electronic applications. It is highly abundant in nature in the form of crystalline quartz and can be grown extremely pure by experimental techniques. 
A prominent defect is the neutral [AlO$_4$]$^0$ center, which has been identified by its electron paramagnetic resonance (EPR) signature\cite{epr-1955,obrian-1955,gtensor-exp1,gtensor-exp2,weil-1980,weil-1981}. Here, an Al$^{3+}$-ion substitutes a Si-ion, leaving an unpaired electron at one of the four oxygen atoms adjacent to the Al center. The corresponding localized spin gives rise to an ESR signal. This defect is believed to cause the smoky coloration of quartz crystals\cite{obrian-1955,schirmer-1976}.
On theoretical level, this 'classical' model has been confirmed by cluster calculations on unrestricted Hartree-Fock or hybrid functional level\cite{adrian-1985,sim-1991,pacchioni-2000,to-2005} 
, yielding a polaronic hole localization and hyperfine coupling constants with $^{27}$Al, $^{17}$O and $^{29}$Si in good agreement with experiments. It was shown that the defect undergoes local symmetry breaking and Jahn-Teller reconstruction with one oxygen atom relaxing away from the Al ion. 
At the other hand, density functional theory (DFT) calculations predicted that the hole and spin are delocalized over all four oxygen atoms and no symmetry breaking between the oxygen sites\cite{continenza-1996,laegsgard-2000}. Similar predictions occured for other oxide wide-gap semiconductors, where DFT on the level of the local density (LDA) or generalized gradient approximation (GGA) level fails to correctly predict the polaronic hole localization on oxygen, such as in case of cation vacancies in ZnO\cite{chan-2009,clark-ZnO,cavalho-ZnO}, HfO$_2$\cite{clark-hfo2}, TiO$_2$\cite{deak-tio2} and acceptor impurities in GaN\cite{lany-GaN}, In$_2$O$_3$ or Sn$_2$O$_3$\cite{lany-2009}. The observed delocalization of the hole arises from the residual one-electron self-interaction from the Hartree energy, which is insufficiently canceled by the exchange-correlation term of LDA and GGA type functionals, thus promoting artificial delocalization\cite{cohen-2008}.\\
Similarly, the self-interaction also contributes to the underestimation of experimental band gaps by predicting valence (conduction) bands at too high (low) energies. 
Indeed, Mauri \emph{et al.}\cite{mauri-2005} showed that self-interaction corrected LDA\cite{sic-lda} calculations, where the contributions due to self-interaction are explicitly substracted in the energy expression, favor a distorted geometry and trapped hole. On the other hand, Hartree-Fock exchange completely cancels self-interaction contributions and inclusion of HF exchange in the exchange-correlation functional typically can compensate for the shortcomings of LDA and GGA. However, pure HF results in a gross over-estimate of the band gap. 
It is interesting that some of these hybrid functionals, such as B3LYP, do not give the full hole localization in SiO2\cite{pacchioni-2000,laegsgard-2001}. This suggests that a minimum amount of exact exchange is needed to give this result correctly\cite{laegsgard-2001}.\\
In this paper, we investigate the [AlO$_4$]$^0$ center using periodic boundary conditions and the screened-exchange (sX-LDA) hybrid functional, which includes a screened version of the full exact exchange. It is essentially self-interaction free for all electron spacings less than the screening length and improves on the predicted band gaps\cite{byklein-SX,Seidl-SX,clark-2010}. We show that sX-LDA can restore the localization of the polaronic hole on one oxygen atom. We further provide calculated EPR parameters, which we find to be in good agreement with experiment. To the best of our knowledge, calculated Land\'e g tensors of the [AlO$_4$]$^0$ center have not been reported so far.

\section{Method}
The calculations were performed in the frame of spin-polarized density functional theory (DFT) using the hybrid functional sX-LDA\cite{byklein-SX,Seidl-SX}, which has been recently\cite{clark-sx} implemented in the planewave code CASTEP\cite{castep}. Here, the self-energy of an electron in the crystal is approximated by a combination of a short-range Hartree-Fock exchange-type term
\begin{eqnarray}
\rho_{ij}(\mbox{\bf{r}})&=&\phi_i^*(\mbox{\bf{r}})\phi_j(\mbox{\bf{r}})\nonumber\\
E^{sX}_x[{\phi}] &\sim& \sum_{i,j}^{occ}\iint \frac{\rho_{ij}(\mbox{\bf{r}})\phi_j(\mbox{\bf{r}})\e^{-k_s|\bf{r}-\bf{r}'|}\rho^*_{ij}(\mbox{\bf{r'}})}{|\bf{r}-\bf{r}'|}d^3r'd^3r\nonumber
\end{eqnarray}
and a long range, local density dependent, LDA-type term
\begin{equation*}
E^{\mbox{\tiny LDA,LR}}_{xc}[\rho] = E^{\mbox{\tiny LDA}}_{\mbox{\tiny xc}}[\rho] - E^{\mbox{\tiny sX,loc}}_{\mbox{\tiny x}}[\rho].
\end{equation*}
This range-separation into short and long range terms is similar to that in other functionals like the Heyd-Scuseria-Ernzerhof (HSE) version\cite{hse03,hse06}.
In theory, the Thomas-Fermi wave vector $k_{TF}$, which depends on the average charge density, is used as the inverse screening length $k_s$. In practice, the TF wavevector needs to be chosen carefully in terms of the density of valence electrons. Here we use a fixed value of 0.76\,bohr, which works well for sp semiconductors.\\
The calculations were done in two steps: In the first step, we used periodic boundary conditions to model the defect in the solid by a 2x2x2 supercell of the common (hexagonal) 9 atom unit cell of $\alpha$-SiO2 with an Al atoms substituting one of the 24 Si atoms, and optimized the atomic positions while keeping the lattice constants at the experimental values. The integrals in reciprocal space were approximated by the values at the Baldereschi k-point for hexagonal lattices\cite{baldereschi} and we used OPIUM\cite{opium1,*opium2} pseudopotentials with a cutoff energy of 800 eV to model the valence electrons of Si and O.\\
In the second step, we calculated the electron paramagnetic resonance (EPR) parameters with the Quantum Chemistry code ORCA\cite{orca1,*orca2}, which particularly aims at the spectroscopic properties of open-shell molecules. We used a 33 atoms cluster, which we obtained by cutting out the first three atomic shells surrounding the Al center from the previously relaxed 72 atom supercell and passivating the dangling bonds of the outmost 12 oxygen atoms by hydrogen. The atoms were represented by all-electron def2-TZVP sets from the Karlsruhe group\cite{def2-TZP}. The exception are the oxygen atoms directly adjacent to the Al-center that were described by Barone's EPR-III\cite{EPRIII} basis set, which was specifically designed for the calculation of EPR properties. As ORCA does not offer screened hybrid functionals, we chose to use to use Becke's 'Half and Half' (BHandHLYP)\cite{BHandHLYP} hybrid functional
\begin{equation*}
E_{\mbox{\tiny xc}}^{\mbox{\tiny BHandHLYP}}[{\psi}] = \frac{1}{2}(E_{\mbox{\tiny x}}^{\mbox{\tiny HF}}[\psi] + E_{\mbox{\tiny x}}^{\mbox{\tiny B88}}[n]) + E_{\mbox{\tiny c}}^{\mbox{\tiny LYP}}[n]
\end{equation*}
instead.

\section{Results and Discussion}

\begin{figure}[tbh]
\centering
\includegraphics*[width=0.98\columnwidth]{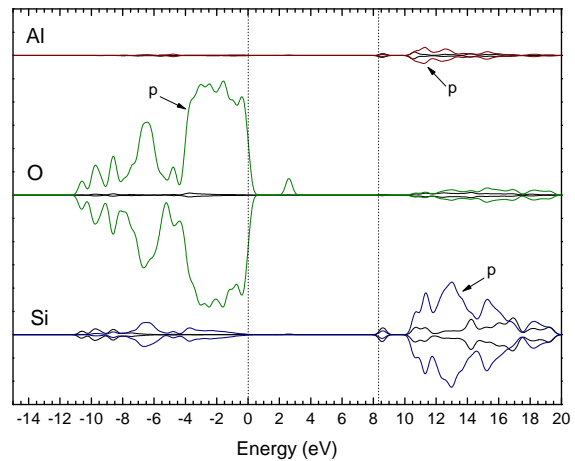}
\caption{\label{fig:SiO2-pdos} (Color online) Partial density of states of a 72 atom supercell of SiO2:Al from sX-LDA calculations. The dashed lines represent the energy of the valence band maximum and the conduction band minimum.}
\end{figure}

\begin{figure}[tbh]
\centering
\includegraphics*[width=0.95\columnwidth]{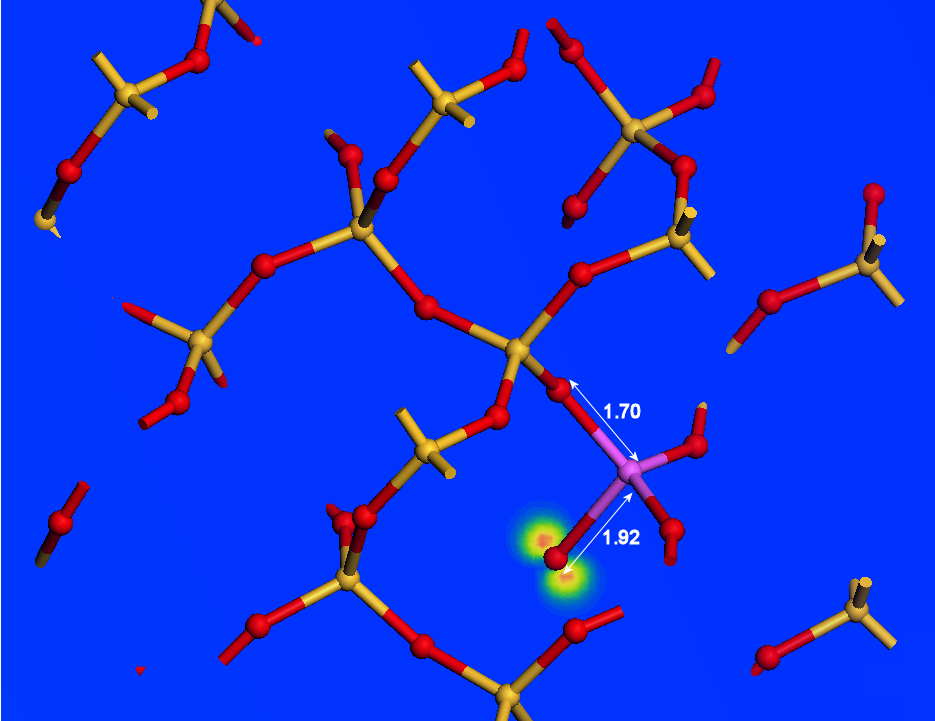}
\caption{\label{fig:SiO2-defect} (Color online) Corresponding plot of the spin density of SiO2:Al around the Al center. The system undergoes a polaronic distortion and the spin is localized almost exclusively at one of the four oxygen atoms adjacent to the Al impurity. Some atoms are hidden by the spin density isoplane.}
\end{figure}

\begin{table*}[tb]

\centering
\caption{\label{tab:SiO2-EPR} Calculated principle values of the g-tensor and the hyperfine matrix of the [AlO$_4$]$^0$ center from calculations on a Al(O$_4$SiH$_3$)$_4$ cluster, and comparisons with previous theoretical from other authors and experiments. Following Ref.~\cite{to-2005}, the oxygen labels O$^*$ and O$^{1,2,3}$ refer to the oxygen atom with the hole and the three remaining oxygen atoms adjacent to the aluminium center, respectively. The values for $^{29}$Si are given for the Si nucleus with he largest contribution.}
\begin{ruledtabular}
\begin{tabular*}{\columnwidth}{@{\extracolsep{\fill}} c  c  c c c c }
\hline
&&This work&UHF\cite{pacchioni-2000}\footnotemark[1]&BB1K\cite{to-2005}\footnotemark[2]&Exp\cite{gtensor-exp2,weil-1980,weil-1981}\\
\hline
Spin&O$^*$&0.93&1.04&0.81&\\
(in $\mu_B$)&O$^{1,2,3}$&0.01-0.05&<0.01&0.00-0.20&\\
&&&&\\
&g$_1$&2.0031&&&2.0024$\pm$0.0003\\
$g$ tensor&g$_2$&2.0093&&&2.0091$\pm$0.0003\\
&g$_3$&2.0412&&&2.0614$\pm$0.0003\\
&g$_{\mbox{\tiny iso}}$&2.0178699&&&\\
&&&&\\
$^{17}$O$^*$&A$_1$&-119.225&-128.6&-109.46&-111.0\\
hyperfine&A$_2$&22.544&11.6&23.48&15.2\\
coupling&A$_3$&22.86&13.6&23.76&17.8\\
matrix&A$_{\mbox{\tiny iso}}$&-24.607&-34.5&-20.74&-26.0\\
parameters&B$_1$&-94.618&-94.1&-88.72&-85.0\\
(in G)&B$_2$&47.151&46.1&44.22&41.2\\
&B$_3$&47.467&48.1&44.50&43.8\\
&&&&\\
$^{27}$Al&A$_1$&-6.236&-5.1&-7.09&-6.2\\
hyperfine&A$_2$&-5.972&-5.1&-6.99&-6.1\\
coupling&A$_3$&-4.953&-4.2&-5.90&-5.1\\
matrix&A$_{\mbox{\tiny iso}}$&-5.721&-4.8&-6.66&-5.8\\
parameters&B$_1$&-0.516&-0.3&-0.43&-0.4\\
(in G)&B$_2$&-0.251&-0.3&-0.33&-0.3\\
&B$_3$&0.767&0.6&0.76&0.7\\
&&&&\\
$^{29}$Si&A$_1$&8.744&15.5&10.11&10.8\\
hyperfine&A$_2$&9.347&17.4&10.57&11.4\\
coupling&A$_3$&9.524&17.8&10.77&11.6\\
matrix&A$_{\mbox{\tiny iso}}$&9.205&16.9&10.48&11.3\\
parameters&B$_1$&-0.403&-1.4&-0.37&-0.5\\
(in G)&B$_2$&0.142&0.5&0.09&0.1\\
&B$_3$&0.319&0.9&0.29&0.3\\
\hline
\end{tabular*}
\footnotetext[1]{Obtained with an Al(OSiH$_3$)$_4$ cluster.}
\footnotetext[2]{Obtained with an AlO$_{16}$Si$_{16}$H$_{36}$ cluster.}
\end{ruledtabular}
\end{table*}

Fig.~\ref{fig:SiO2-pdos} shows the calculated partial density of states (PDOS) for a 72 atom supercell of $\alpha$-SiO$_2$ containing a single Al center. Clearly, the impurity introduces a single spin-polarized defect state at an energy of 2.5\,eV above the valence band maximum. The defect state is predominantly of O $2p$ character with small contributions from O $2s$ and Si $3p$ mixed in. We find that the overall magnetic moment of 0.99\,$\mu_B$ of the system is localized almost exclusively at one single oxygen atom, refer to the spin density plot in Fig.~\ref{fig:SiO2-defect}. Together with the localized hole, our sX-LDA approach yields an asymmetric distortion of the geometry surrounding the Al atom. We find an Al-O bond length of 1.92\,\AA\space for the oxygen atom carrying the localized hole, corresponding to an bond elongation of ~13\% compared to the other three Al-O bonds (1.69-1.71\AA). This prediction is in reasonable agreement with estimation from EPR measurements\cite{weil-1981}(12\% bond elongation).\\

We note that the degree of spin localization of depends on the included portion of Hartree-Fock exchange in the functional. Pacchioni \emph{et al.}\cite{pacchioni-2000} reported that the popular B3LYP functional (20\% HF exchange) predicts a partial spin localization with spin populations of 0.29 on two oxygen atoms and 0.21 on the remaining two. Correspondingly, the Al-O bond lengths in this case are very similar. On the other hand, To \emph{et al.}\cite{to-2005} found a bond elongation of 12\% using the BB1K functional (42\% HF exchange), whereas Pacchioni \emph{et al.}\cite{pacchioni-2000} reported a 14\% elongation from their 100\% Hartree-Fock calculations. Our cluster calculations using the BHandHLYP hybrid functional (50\% exact exchange) yield spin localization and defect geometry very similar to our sX-LDA results.\\

In the next step, we calculated the EPR parameters for the defect. The EPR spectrum can be modeled by an Hamiltonian
\begin{equation}
H_{eff} = \frac{\alpha}{2}\mathbf{S}\cdot\mathbf{g}\cdot\mathbf{B} + \sum_i\mathbf{S}\cdot\mathbf{A}\cdot\mathbf{I}_i\label{eq:Heff}
\end{equation}
where $\alpha$ is the fine structure constant. The first term of Eq.~\ref{eq:Heff} describes the coupling of the spin momenta $\mathbf{S}$ of the unpaired electrons with an external magnetic field $\mathbf{B}$ by the $g$ tensor 
\begin{equation*}
\mathbf{g}=
 \begin{pmatrix}
  g_{1} & 0 & 0\\
  0 & g_{2} & 0\\
  0 & 0 & g_{3}
 \end{pmatrix}
\end{equation*}
The second term of Eq.~\ref{eq:Heff} represents the hyperfine coupling of the electron spin with the nuclear spins can be described in terms of the hyperfine matrix $\mathbf{A}$, which can be separated into its isotropic and its anisotropic part $\mathbf{B}$ (related to dipolar interaction)
\begin{equation*}
\mathbf{A}=
 \begin{pmatrix}
  A_{1} & 0 & 0\\
  0 & A_{2} & 0\\
  0 & 0 & A_{3}
 \end{pmatrix}
+
A_{\mbox{\tiny iso}}\mathbf{I}+
 \begin{pmatrix}
  B_{1} & 0 & 0\\
  0 & B_{2} & 0\\
  0 & 0 & B_{3}
 \end{pmatrix}
.
\end{equation*}
A problem arises from the fact that the $g$ tensor is not gauge invariant, so that the results technically depend on the choice of origin in the calculation. While the origin dependence is usually small as long as sufficiently large basis sets are used, we have checked the sensitivity of the results with respect to the origin. We found that on GGA level the center of the electron charge gives practically the same values as calculations using the fully invariant IGLO (individual gauges for localized orbitals)\cite{IGLO-paper} procedure (which, however, is not rigorous for hybrid functionals) and used this point as the origin for our calculations. Table~\ref{tab:SiO2-EPR} shows the obtained principle values of the g-tensor and the hyperfine matrix in comparison with previous theoretical results and experiment. In case of the g-tensor, our computed values for the two smaller principle values $g_{\mbox{\tiny 1}}$=2.0031, $g_{\mbox{\tiny 2}}$=2.0093 are in very good agreement with the experimental estimations $g_{\mbox{\tiny 1,e}}$=2.0024 and $g_{\mbox{\tiny 2,e}}$=2.0091. The deviations for the third principal value are larger. We find $g_{\mbox{\tiny 3}}$=2.0412, which is considerably smaller than $g_{\mbox{\tiny 3,e}}$=2.0614 as reported by Schnadt \emph{et al}\cite{gtensor-exp2}\\ 
For the hyperfine coupling, we find a good prediction of our approach for the investigated $^{17}$O, $^{27}$Al and $^{29}Si$ nucleii. This is particularly true for $^{27}$Al, where our results, being the golden mean of the reported values of UHF\cite{pacchioni-2000} and BB1K\cite{to-2005} calculations, are remarkably close to experiment. For $^{29}$Si, our calculations slightly underestimate the experimental values and yield similar results to those from To \emph{et al}. This is entirely due to the underestimation of the isotropic component of $\mathbf{A}$, A$_{\mbox{\tiny iso}}$ by 2\,G, while the (small) dipolar interaction is reproduced accurately. The dominant contribution to hyperfine coupling is to be expected from the oxygen atom with the localized hole. Here, the electron-nucleus interaction contains a strong dipolar component and the experimentally obtained hyperfine matrix shows a strong anisotropy for the three principal axes. The UHF calculations by Pacchioni \emph{et al.} underestimated the experimental values $A_{1,e}$=-111.00\,G, $A_{2,e}1$=15.2\,G and $A_{3,e}$=17.8\,G due to a too large isotropic contribution. In contrast, the BB1K calculations predicted a considerably weaker A$_{\mbox{\tiny iso}}$, resulting in a slight underestimation of $A_{1}$ and overestimation of $A_{2}$ and $A_{3}$ and the opposite behavior for the anisotropic matrix. We find the hyperfine parameters $A_1$=-119.22\,G, $A_2$=22.54\,G and $A_3$=22.86\,G, which overestimate all three of the experimental values. The reason is a too strong anisotropic component in our hyperfine matrix, as Tab~\ref{tab:SiO2-EPR} shows. At the other hand, our predicted A$_{\mbox{\tiny iso}}$ is favorably close to the experimental value and a better prediction than the values from the other two methods. In fact, the differences between our parameters and those from Pacchioni \emph{et al.} arise from the different isotropic contributions, while the dipolar interaction is close to identical in both cases. The dipolar contribution depends on the angular momentum of the unpaired electron and hence is roughly proportional to the population of the O 2$p$ orbital and the localization of the spin density. To \emph{et al} have reported that increasing the size of the cluster in their case led to a stronger spin localization and a larger dipolar interaction. We could thus argue that our calculations slightly overestimate the localization of the electron and the correct spin population of the oxygen atom is about 0.9\,$\mu_B$. This is in line with the observation that overall the slightly smaller portion of Hartree-Fock exchange in the BB1K functional compared to BHandHLYP seems to benefit the prediction of hyperfine parameters for all investigated nucleii.

\section{Conclusion}
We showed that sufficient inclusion of Hartree-Fock in hybrid functionals does improve on DFT supercell calculations and can correctly describe the polaronic hole and the corresponding symmetry distortion of the neutral Al impurity in $\alpha$-quartz. The observed localization is inherently connected to the reduced self-interaction in hybrid functionals, which shows in by its dependence on the ratio of HF:DFT exchange. The validity presented approach was further shown by a subsequent calculation of the g-tensor and the hyperfine coupling matrix using a cluster approximation for the sX-LDA geometry, which was in good agreement with the conclusions from previous EPR measurements.



%

\end{document}